-------------------------------------------------------------------------------------------------------------------

**Title**

**Reference Class Forecasting for Hong Kong's major roadworks projects**

Author 1

**Prof Bent Flyvbjerg**, Dr Techn., Dr Scient., PhD

First BT Professor and founding Chair of Major Programme Management,

Saïd Business School, University of Oxford

Author 2

**Ir CK Hon**, JP, BSc(Eng), MPA, CEng, FHKIE, FICE, FCIArb

Permanent Secretary for Development (Works), the Government of Hong Kong SAR

Author 3

**Ir Dr WH Fok**, MSc, MA, EngD, CEng, FHKIE, FICE, FIStructE, FCIHT, FCIArb, FHKIArb, APMP

Director Projects, Advisian Limited

**Full contact details of corresponding author**

Ir Dr WH Fok

Advisian Limited

A:  32/F, Sunlight Tower, 248 Queen's Road East, Wanchai, Hong Kong

T:  (852) 3556 7356

M:  (852) 9373 5775

E:  wh.fok@advisian.com

**Abstract**

Reference Class Forecasting (RCF) is a method to remove optimism bias and strategic misrepresentation in cost and time to completion forecasting of projects and programmes.

In September 2012, the Development Bureau of the Government of the Hong Kong Special Administrative Region commissioned a study to test the feasibility of using RCF in Hong Kong, the first of its kind application in the Asia-Pacific region.

This study comprises 25 roadwork projects. For these projects cost and time to completion forecast and actual data were retrieved. The analysis established and verified the statistical distribution of the forecast accuracy at various stages of project development and benchmarked the projects against a sample of 863 similar projects.

The study contributed to understand how to improve forecasts by (1) de-biasing early estimates, (2) explicitly consider the risk appetite of decision makers, and (3) safeguard public funding allocation by balancing exceedance and under-use of project budgets.

# 1. Introduction

Investments in public infrastructure have historically been seen as a key policy to stimulate economic growth. Calderon, Serven, and Moral-Benito (2011) estimated that a 10% increase in infrastructure provision increases long-term economic output by 1%-5%. A recent report by the McKinsey Global Institute (2013) argued that deferring infrastructure investments stifles growth and recovery. The report also estimates that the current infrastructure stock is on average 70% of a country's GDP (China 76%), which is seen as indication that infrastructure spending needs to grow with countries' GDP growth targets. In the case of China, the projected GDP growth by 2013-30 would translate into growing the needed investment into infrastructure from 6.4% of the GDP to 8.5% of the GDP. A historic analysis by the OECD/ITF (2013) showed that countries in Western Europe, North America, and Australia used investments into road infrastructure, although while reducing maintenance spend, to stimulate economic growth during the recession 2008-2011. However, research by Flyvbjerg, Bruzelius, and Rothengatter (2003) has shown that capital investment projects often experience cost overruns and completion delays and subsequent research questions the positive contribution of infrastructure megaprojects (Flyvbjerg et al. 2003),

*1.1 The Reference Class Forecasting (RCF) method*

Reference class forecasting is the method of predicting the future, through looking at similar past situations and their outcomes. Reference class forecasting predicts the outcome of a planned action based on actual outcomes in a reference class of similar actions to that being forecasted. The theories behind reference class forecasting were developed by Daniel Kahneman and Amos Tversky. The theoretical work helped Kahneman win the Nobel Prize in Economics.

Kahneman and Tversky (1979) found that human judgment is generally optimistic due to overconfidence and insufficient consideration of distributional information about outcomes. Therefore, people tend to underestimate the costs, completion times, and risks of planned actions, whereas they tend to overestimate the benefits of those same actions. Such error is caused by actors taking an "inside view," where focus is on the constituents of the specific planned action instead of on the actual outcomes of similar ventures that have already been completed.

Kahneman and Tversky (1979) concluded that disregard of distributional information, that is a focus on single point estimates is perhaps the major source of error in forecasting. On that basis they recommended that forecasters "should therefore make every effort to frame the forecasting problem so as to facilitate utilising all the distributional information that is available". In other words, forecasts should not be confined to the most likely estimate but present the full range of estimates. While, newer methods such as Monte Carlo Simulations attempt to present a range of outcomes, these simulations are still the result of an inside view and therefore prone to underestimation. Kahneman and Tversky (1979) further argue that using distributional information from previous ventures similar to the one being forecast is called taking an "outside view". Reference class forecasting is a method for taking an outside view on planned actions.

RCF for a specific project involves the following three steps:
1. Identify a reference class of past, similar projects.
2. Establish a probability distribution for the selected reference class for the parameter that is being forecast.
3. Compare the specific project with the reference class distribution, in order to establish the most likely outcome for the specific project.

The key benefit of RCF is that it replaces assumptions, e.g. assumed adverse events, assumed uncertainty in estimates, with data. It should be noted that RCF assumes that a project performs no better or worse than past similar projects. On this basis adjustments can be made to the Reference Class Forecast. However it should be stressed that adjustments might re-introduce the bias of assumptions back into the forecast and should only be made based on strong evidence that the project is indeed better or worse than past projects.

*1.2 The use of RCF in cost and other forecasting of major projects*
RCF has been used by the UK Department for Transport since 2004 to implement the UK Treasury's Green Book Guidance on Optimism Bias Uplifts. It has since been used on all major UK transport infrastructure projects and programmes. It has also been used outside the transport sector in a broad variety of projects and programmes across departments ranging from the Government's review of public private partnerships, the renewal of the British School Estate, equipment purchases in the UK health care system to the turn-around programme of Rover.

RCF has also developed into a method used globally. In 2005, the American Planning Association endorsed RCF. Similarly, the Project Management Institute has included the concept of taking the 'outside view' in its standard for project cost estimation. Like the UK, Denmark has made RCF mandatory for large rail and road projects. Furthermore, RCF has been used on individual projects by the governments of Sweden, Switzerland, Norway, the Netherlands, South Africa and Australia, among others.

RCF was pivotal in the US Government Accountability Office's assessment of the business case of California High Speed Rail and is currently being used in Europe's largest civil engineering projects, Crossrail and HS2.

Daniel Kahneman has called RCF as developed by Flyvbjerg and COWI (2004) "the single most important piece of advice regarding how to increase accuracy in forecasting through improved methods." (Kahneman 2011, p. 251)

*1.3 The benefits of using RCF*
The UK experience of using RCF for more than 10 years has shown that due to the method a number of projects have been cancelled. Projects that were implemented after RCF had been applied have shown reduced cost overruns in comparison to projects that have not been de-biased. "We are tending now to find that with experience, our project costs remain very much under control and the optimism bias added in appraisals is too high against the actual delivery," observed a UK DfT official (personal communication December 2012).

Discussing the value of RCF with project managers, who have applied the method, we learned that the value of RCF becomes particularly visible in the latter half of a project. When project budgets are exhausted and contingencies spent, which is common for projects without RCF, project managers spend most of their time on renegotiating project scope, raising new funding, reducing the cost of the outstanding works, and answering critique of bad project management in the media. Typically this point is reached in the latter half of project delivery. However, anecdotal feedback from project managers suggests that projects with an un-biased level of contingencies, allow project managers to focus on what is really important: completing the project.

## 2. The First RCF Study in Hong Kong and Asia-Pacific

The idea of RCF was first introduced to Hong Kong in late 2010. In September 2012, a study was commissioned by the Development Bureau (2012) to study the feasibility of using RCF in Hong Kong, with the major roadworks projects of the Highways Department as a pilot reference class. It involved the retrieval of cost and completion time information of 25 completed projects for statistical treatment such that a reference class on cost and completion time forecasting of roadworks projects could be established. The Study was the first of its kind in the Asia-Pacific region.

*2.1 The development process of Hong Kong's public works projects*

All public works projects including roadworks projects are developed under well-established government procedures. According to Chapter 2 of the Project Administration Handbook for Civil Engineering Works issued by the Government of the HKSAR (2014), or its earlier versions, public works projects go through various stages of development before construction as categorised below:

- Category C: projects that have been broadly accepted on the basis of a Project Definition Statement and Technical Feasibility Statement that has been approved by the Works Branch of the Development Bureau
- Category B: projects which have had resources earmarked internally, but they are not yet presented to Finance Committee of the Legislative Council for approval and they cannot incur expenditure other than on pre-construction works including investigation and design
- Category A: projects which are ready in all respects for tenders to be invited and construction works to proceed, and have been granted an approved project estimate by Finance Committee

When a project is being upgraded to each of these categories, an estimate of the cost prepared by the respective works department will be logged into the Capital Works Programme of the government. It is natural that the cost estimate of a project at Category A is much closer to the final cost of the project than that at Category C, since more details of the project have been worked out when a project progresses from Category C to Category B, then to Category A.

*2.2 The current practice of cost forecasting and the potential use of RCF*

When preparing for the cost estimate of a project at each category stage, the works department will produce the base estimate based on quantities of the latest project plan and the most appropriate rates.

Since mid-1993, the assessment of the contingencies of a project follows the standard process called Estimating Using Risk Analysis (ERA) as directed by the technical circular issued by the Works Branch (1993), then a major branch under the Government Secretariat. In the process, risk items of a project are identified and quantified based on experience and knowledge of project officers on similar projects. The summation of all risk values becomes the contingency value to be added to the base estimate of the project to become its project estimate for logging into the Capital Works Programme. The same process is applied when a project upgrades to Category C, Category B and Category A status.

The ERA process is by nature a bottom-up process based on the "inside view" of the project officers to determine the risk values of a certain project. In contrast, the RCF provides a data-driven "outside view" of the risk values of a certain project. First, RCF establishes the full distributional information of the risk exposure of a project. Second, RCF provides an uplift factor to the base estimate to become the project estimate, based on the risk appetite of the decision maker. This is regarded as a top-down process since the uplift factor corresponding to the decision makers' acceptable risk of budget overrun is derived from the statistics of comparison between the constant price, ie same time value, of the final project cost and the base estimate of a reference class of similar past projects, hence an "outside view". Both, ERA and RCF, are methods of quantitative risk analysis based on the risk-free base estimate. RCF offers an "outside view" to scrutinise the "inside view" ERA results and decision makers can subsequently allocate contingencies corresponding to their risk appetite.

## 3. The study process

For this study data was collected by first identifying a list of past projects each larger than HKD100 million in final outturn cost, a criteria set after discussing with the Highways Department about what is considered a large-scale project. The study aimed to collect data of 25 projects, which was a trade off to between the statistical validity of the results and the effort needed to collect the archival data. Data on the most recently completed 25 projects above HKD100 million was collected. The data include the scope and timeline of each project as well as base estimates and contingency estimates at the stages of upgrading to Category C, Category B and Category A as well as the final cost. All these projects are major roadworks projects constructed between 1993-2011. The earliest upgrade to Category C was in 1986.

This approach to data collection was taken to ensure a representative view of the projects undertaken by the Highways Department. The availability of archival records, one could argue, might have an impact on the performance that a study will find if badly performing projects fail to document their plans and progress reports. Thus we have previously argued, e.g. in the international benchmarking, that the results err on the conservative side.

*3.1 Retrieval and adjustment of project data*

Data retrieved for the different category stages were verified by cross-checking with available detailed breakdown of the figures of the respective project, such as checking the consistency of project estimate with the base estimate plus contingencies at the same category stage.

For the 25 projects cost and time estimates were available for 23 projects at Category C, 22 projects at Category B, and 20 projects at Category A. All estimates and final outturn cost were then adjusted to the same year level using the public works price deflators issued by the Government of the Hong Kong Special Administrative Region.

For the projects where actual disbursement record of projects could not be retrieved, statistically established disbursement profiles in Table 1 are assumed. The assumption of a distribution profile is commonly made to convert year-of-expenditure cost into constant cost, i.e. to remove the effects of inflation (see for example Bacon and Besant-Jones 1996, Ansar et al. 2014). For instance, a 3-year project is spending 17% of the total outturn cost in year 1, 65% in year 2, and 18% in year 3. A 5 year project spends 6% in year 1, 22% in year 2, 43% in year 3 and so on.

| Project length in years | Disbursement Profile (as % of total outturn cost in Xth year of the project) | | | | | | | | | |
|---|---|---|---|---|---|---|---|---|---|---|
| | 1st year | 2nd year | 3rd year | 4th year | 5th year | 6th year | 7th year | 8th year | 9th year | 10th year |
| 1 | 100% | | | | | | | | | |
| 2 | 49% | 51% | | | | | | | | |
| 3 | 17% | 65% | 18% | | | | | | | |
| 4 | 9% | 40% | 42% | 10% | | | | | | |
| 5 | 6% | 22% | 43% | 23% | 6% | | | | | |
| 6 | 4% | 13% | 32% | 33% | 14% | 5% | | | | |
| 7 | 3% | 8% | 21% | 32% | 23% | 9% | 4% | | | |
| 8 | 3% | 4% | 10% | 20% | 25% | 20% | 11% | 7% | | |
| 9 | 3% | 4% | 10% | 20% | 25% | 20% | 11% | 4% | 3% | |
| 10 | 2% | 3% | 7% | 14% | 21% | 22% | 15% | 8% | 4% | 3% |

Table 1: Disbursement Profiles

An analysis of the adjusted data revealed statistically significant improvements in forecasting accuracy for different vintages of estimates (cf. Figure 1). This improvement in forecasting accuracy was associated with a more realistic estimation of the base cost, which was reflected in increased unit cost estimates in real terms. The analysis over time also established a shift in estimation accuracy after mid-1993, when the ERA was introduced. The shift from before 1993 to after 1993 as shown in Figure 1 is nearly statistically significant even with a small data sample. In order to ensure that the established reference class is relevant to current planning practices, projects started before 1993, i.e., pre-ERA, were excluded from formulating the reference class.

[Figure 1]

The cost overrun is defined as the actual minus estimated costs devided by the estimated cost and expressed in percentages. Cost overruns are calculated based on constant prices. Schedule overrun is the difference between actual time to completion and the estimated time to completion divided by the estimated time to completion. Estimated time to completion is defined as duration of a project from the date of the upgrade to Category C to the planned completion date; while actual time to completion is defined as duration from the date of the upgrade to Category C to the date of actual completion. Time to completion data were also retrieved while screening through the project files. Time to completion estimate were available for 22 projects at Category C, 23 projects at Category B and 25 projects at Category A.

### 3.2 Formation of uplift curves of the reference class

With all the valid cost and time data, the reference class with the uplifts for cost and time to completion estimates at the upgrade to Category C, Category B, and Category A were established.

First, the cost and time to completion overruns at the decision points were calculated. Second, the uplifts $Q$ were established by identifying $Q(p) = \inf\{x: p \leq PR(X \leq x)\}$, where $x$ are the overruns, $p$ is the probability of a cost overrun and $X$ is a given value of $x$. In other words, for all probabilities between 0 and 1 we established the maximum overrun that was not exceeded in the historical data. The uplifts are thus comparable to the risk exposure levels derived from Monte Carlo simulations, which are commonly used in quantitative risk assessments and which also identify the risk exposure levels through examining the quantiles of the distribution.

The cost or time to completion uplifts at each category stage is represented by a curve similar to Figure 2 and Figure 3. The figures Loess Curves, i.e. smoothed local regressions, given the uplifts in cost and time to completion, respectively, for the upgrade to Category C.

[Figure 2 & Figure 3]

Two levels of uplifts that correspond to the risk levels that project owners are typically willing to take were highlighted. In large portfolios where cost overruns in one project can be offset by cost underruns in other projects most organisations accept a 50% chance of cost overruns after adding contingencies. In other words, they aim for 50% certainty of the revised estimate (P50, for portfolio management). When forecasting individual projects or projects with a high potential impact on budgets, organisations typically accept less risk of overruns, e.g., only 20%, and thus choose a higher level of certainty, e.g. 80% (P80, for management of individual projects). Using Figure 2 as an example, at a 20% acceptable chance of overruns the data suggest an uplift of +44%, while at a 50% acceptable chance of overruns the data suggest an uplift of +13%. The uplifts are based on the quantity estimate that is the baseline cost in constant HKD terms, excluding all contingencies added during the ERA and excluding inflation.

### 3.3 Validation of the reference class

To test the robustness of the reference class, the "leave-one-out validation" method was used. The validation works in three steps:

    Step 1:    exclude one project from the data set,
    Step 2:    construct a reference class based on the remaining projects, and
    Step 3:    check whether the suggested uplift would have provided a risk envelope large enough for the excluded project.
    Repeat steps 1-3 for the next project from the data set.

Table 2 shows the results of the validation using Category C baseline cost uplifts data available for 18 projects. First, the project with ID 6736 was excluded from the dataset. Second, using the remaining 17 projects a reference class was constructed. The reference class suggested a +18% P50 uplift and a +46% uplift for P80. Third, the actual cost performance of project 6736 was -47%, which means that both the P50 and the P80 envelope would have provided a sufficient level of contingencies for the project.

| Project No. | P50 uplift | P80 uplift | Actual Cost Performance | Would P50 have prevented overrun? | Would P80 have prevented overrun? |
|---|---|---|---|---|---|
| 6736 | +18% | +46% | -47% | ✔ | ✔ |
| 6757 | +14% | +41% | +47% | ✘ | ✘ |
| 6365 | +14% | +46% | +32% | ✘ | ✔ |
| 6553 | +14% | +41% | +62% | ✘ | ✘ |
| 6580 | +18% | +46% | -37% | ✔ | ✔ |
| 6718 | +18% | +46% | -33% | ✔ | ✔ |
| 6731 | +18% | +46% | -32% | ✔ | ✔ |
| 6759 | +18% | +46% | -7% | ✔ | ✔ |
| 6384 | +14% | +41% | +52% | ✘ | ✘ |
| 642 | +18% | +46% | +14% | ✔ | ✔ |
| 6577 | +18% | +46% | +1% | ✔ | ✔ |
| 6706 | +18% | +46% | -52% | ✔ | ✔ |
| 6541 | +14% | +41% | +68% | ✘ | ✘ |
| 6721 | +14% | +44% | +43% | ✘ | ✔ |
| 6323 | +14% | +46% | +29% | ✘ | ✔ |
| 6694 | +14% | +46% | +31% | ✘ | ✔ |
| 6695 | +18% | +46% | +1% | ✔ | ✔ |
| 6645 | +14% | +46% | +18% | ✘ | ✔ |

Table 2: Leave-one-out validation of cost uplifts for Category C baseline cost estimates

In sum, the table shows that in 9 out of 18 reference classes the P50 uplift would have been sufficient to prevent a cost overrun. In other words, to validate the reference classes we treat each project as if it is to be de-risked by the uplifts suggested by a reference class based on the other projects. In 50% of these validations the uplift would have been exceeded, which validates that the P50 uplift of this reference class corresponds to a 50% acceptable chance of overruns.

Similarly, in 14 out of 18 references classes, the P80 uplift would have provided a sufficient contingency to prevent a cost overrun. This means in 78% of the cases the risk envelope was big enough. This shows that the P80 uplift corresponds to a 22% acceptable chance of cost overruns, which is close to the targeted risk level of 20%.

Taken together, these results show the reference class to be robust in the sense that the reference class would have been sufficient to prevent cost overruns through the application of uplifts had any of the projects in our sample been currently in planning.

4. **Outcomes of the Study**

*4.1 Uplifts for cost and time to completion forecasting*

Based on the uplift curves in Figure 2 and Figure 3 for Category C and the respective curves for Category B and Category A, the results of the cost and time to completion reference class forecast for the upgrades to each category stage at both a 50% acceptable chance of overruns (P50) and a 20% acceptable chance of overruns (P80) can be summarised in the following table.

|  | Level of certainty | Required uplifts | | |
|---|---|---|---|---|
|  |  | Cat C | Cat B | Cat A |
| Base cost uplift[1] | P50 | +13% | +7% | -1% |
|  | P80 | +44% | +34% | +14% |
| Schedule uplift[2] | P50 | +26% | +22% | +8% |
|  | P80 | +75% | +53% | +26% |

[1]Total estimated budget, excluding inflation and excluding bottom-up contingencies
[2]Measured from the date of the estimate to the date of project completions

Table 3: Summary of the reference class forecast uplifts

The reference class forecasts show that uncertainty reduces over time in the project estimates. At upgrade to Category C the P80 uplift should be +44% for cost and +75% for time to completion, with the upgrade to Category B the uplifts have reduced because of a reduction in uncertainty of forecasts and should be +34% for cost and +53% for time to completion. When the project is upgraded to Category A the appropriate uplifts are +14% for cost and +26% for time to completion.

The results also show that once a project has been upgraded to category A the median budget is quite accurate: the P50 cost uplift is -1%. This shows that the median (P50) project estimate at Category A tends to match with the final cost.

*4.2 International benchmarking*

The available cost and time overruns of Hong Kong's roadworks projects were compared to an international benchmark maintained by researchers working at the University of Oxford. The benchmark database covers 863 road projects from 34 countries. The baseline of the benchmark database is the decision to build, sometimes also called the green light decision. In discussions with project planners in Hong Kong we ascertained that the decision to build corresponded to the upgrade to Category C, because projects that have been upgraded to Category C are rarely abandoned.

|                                      | International Benchmark | Cat C estimates | Cat B estimates | Cat A estimates |
|---|---|---|---|---|
| Average cost overrun                 | +20%       | +11%        | +6%         | -1%         |
| Frequency of cost overruns           | 9 out of 10 | 7 out of 10** | 7 out of 10 | 5 out of 10 |
| Standard deviaton of cost overruns   | 30%        | 38%         | 34%         | 24%         |
| Average schedule overrun             | +38%       | +58%†       | +30%        | +18%        |
| Frequency of schedule overruns       | 6 out of 10 | 8 out of 10* | 8 out of 10 | 9 out of 10 |
| Standard deviaton of schedule overrun| 85%        | 103%        | 39%         | 29%         |
| Average duration (years)             | 5.5        | 8.9***      |             |             |

Note: Only Cat C estimates are directly comparable to the international benchmark
† $p \leq 0.1$
\* $p \leq 0.05$
\*\* $p \leq 0.01$
\*\*\* $p \leq 0.001$

Table 4: International benchmarking of the Hong Kong highways data

As shown in table 4, the roadworks projects in Hong Kong had an average cost overrun of +11%, which is lower than the international benchmark, where the average cost overrun was +20%. This difference, however, is not statistically significant using a non-parametric test. Moreover, the cost overruns were less frequent in Hong Kong: 7 out of 10 Hong Kong road works projects experienced cost overruns, while 9 out of 10 projects in the benchmark database had cost overruns (non-parametric test statistically significant at p = 0.006). This indicates that there is a bias in the Hong Kong data towards cost overruns, i.e. the likelihood of a cost overrun is higher than the likelihood of an underrun. The bias, however, is small than the international benchmark. Moreover, the bias reduces as the estimate progresses towards construction at Category A estimates are as likely to overrun as they are to underrun. Conversely, the standard deviation of the cost overruns in the HK highway projects was 38%, which was greater than 30% of the international benchmark. In other words the HK highway projects showed greater variability in their cost overruns.

Time overruns showed a different pattern compared to the benchmark. The average time to completion overrun of Hong Kong roadworks projects was +58% while the international benchmark was +38% (non-parametric test not statistically significant, p = 0.089). Again the frequency of the time to completion overruns was statistically significantly higher in Hong Kong (8 out of 10 projects) than in the international benchmark (6 out of 10 projects, p = 0.048). This indicates that time estimates are biased towards underestimation, i.e. delays occurring, and unlike the cost estimates this bias does not reduce with estimates closer to the start of construction. Lastly, the variability of time to completion overruns was higher in Hong Kong (103%) than in the international benchmark (85%).

Besides, Hong Kong roadworks projects had a statistically significantly longer average duration of 8.9 years compared to the benchmark of 5.5 years (non-parametric test, $p < 0.001$). A summary of the time spent by the studied projects in the different phases is presented in Figure 6. It shows that generally more than half of the project duration was spent before commencement of the actual construction, indicating longer lead time required for project design and planning, public engagement, etc. in the past two decades.

[Figure 6]

Lastly, the figures showed that both cost and time to completion overruns improved from early estimates to estimates closer to construction. The average cost overruns decreased from +11% at the upgrade to category C to -1% at the upgrade to category A. Similarly, the frequency of cost overruns decreased from 7 out of 10 projects to 5 out of 10 projects. Average time to completion overruns also improved over time from +58% at upgrade to Category C to +18% at upgrade to Category A. In other words the data showed that the accuracy of forecasts increased as the project progresses.

***4.3 Adverse consequences of using uplifts and mitigation through portfolio management***
In this study we argued that RCF shows the most un-biased estimate of the risk exposure of a certain project in its full distribution. For a given risk appetite of the decision maker the distribution suggests a top-down uplift to adjust the bottom-up estimate. The most basic action would be to allocate this uplift to the project as a contingency, in other words to increase the project budget and time. However, there two issues arise with this approach.

First, Parkinson (1958) reckoned that if large contingencies are given to project managers it is highly likely that they will be used up. This phenomenon is well documented in capital investment projects in the public sector. But this study found no strong evidence that this is the case for roadworks projects in Hong Kong, where half the projects in our sample underspent their budget estimate made at the upgrade to Cat C. It also means that not all contingency, which we excluded from the estimates, is used up in the projects in Hong Kong. This unusual pattern might be due to peculiar cultural differences in Hong Kong project management and warrants further research.

Second, as observed in other jurisdictions, e.g. by Liu and Napier (2010), inefficient allocation of funds comes from providing contingencies that are later not spent. A benchmark of the full cost estimate (base cost + contingencies) found evidence of pessimism bias also in Cat C and Cat B cost estimates in this sample.

Thus contingencies need to be managed more intelligently than simply added to project estimates to avoid that (a) contingencies are spent freely and (b) avoid an inefficient allocation of funds. Portfolio management and thus a multi-level governance of project funds can improve the allocation of budgets while tightly controlling the contingency drawdown. A tier contingency scheme could, for example, use the RCF to define contract, project, and portfolio contingencies. In such a governance scheme, contract managers could have a delegated authority over a small contingency, e.g. 10% which is equivalent to approximately P50-P55. The project manager could have delegated authority of another 10% contingency, i.e. up to approximately P60. And at portfolio level an additional contingency of 20% could be held, that covers a total risk exposure up to P80. Such a multi-layered approach to contingency management can ensure that project budgets are not inflated while contingencies are available if and when they are needed.

## 5. Experience gained from the Study

Much efforts of the Study were spent on retrieving and verifying project data. These efforts could have been saved if the relevant data were recorded in a way convenient for the application of the RCF method. A standard proforma was developed as part of the Study so that all cost and time to completion data of future projects that are relevant to the maintenance, updating and application of the reference class can be recorded and retrieved conveniently without much effort. The concerned project officers were trained to record the data into the standard proforma and to use them in updating the uplift curves of the reference class at suitable intervals.

The study has provided valuable insights into how projects are planned and how the maturity of individual schemes increases during the front end process. The study has also shown that early estimates, made when information is scarce and uncertain, can be de-biased to prevent cost overruns and time-to-completion delays using an RCF approach to conservatively bias otherwise optimistic estimates. Moreover, the study has shown that sufficient data exist in Hong Kong and that this data can be used to create value-adding insights for project managers, departments and policy makers. Lastly, the study has highlighted that although RCF alone is a useful method to de-bias project estimates, further enhancements can be made by integrating probabilistic thinking into project incentive schemes and portfolio management regimes.

## 6. Conclusions and recommendations

Countries invest in public infrastructure with a view to increase long-term economic growth. However, capital investment projects often experience cost overruns and time to completion delays. The RCF method is addressing the overrun issue by adopting a top-down approach based on the "outside view" of past similar projects. With the appropriate uplifts to the baseline cost and time to completion estimates, project estimates that correspond to acceptance level of certainty can be worked out. The method has been widely and successfully used in Europe, endorsed by several countries and the first time developed for use ithe Asia-Pacific region.

In the Study, the uplift curves of the reference class of Hong Kong roadworks projects at various development category stages were established by relevant data collected from 25 completed projects. They can be used to determine project estimates with contingencies at the appropriate level of certainty.

On comparing the data with an international benchmark that is equivalent to the baseline cost at Category C stage, Hong Kong performed better in cost forecasting on average. For time to completion forecasting, it was found that generally more than half of project duration was spent before actual construction, i.e. pre-construction stage.

We also argued that the current contingency provisions on project basis might be inefficiently allocated. If the contingencies are pooled across a portfolio of projects, the contingency funds can be used more flexibly and thus more efficiently.

This study offers important contributions to estimation and forecasting practice. We documented three implications:
- (1) How early project estimates can be de-biasing,
- (2) How contingency can be set in a data-driven way that explicitly considers the risk appetite of decision makers, and
- (3) How public funds can be safeguarded, i.e. how exceedance and under-use of project budgets can be balanced through simultaneously governing project contingencies at the project and portfolio level.

This study also contributed to the discussion of reference class forecasting and similar methods by suggesting a method to test the robustness of the reference class forecast and by providing data from the Asia Pacific Region for the very first time.

**Acknowledgements**

The authors would like to acknowledge the Development Bureau and the Highways Department of the Government of the HKSAR for the permission of publishing this paper.

**Figures**

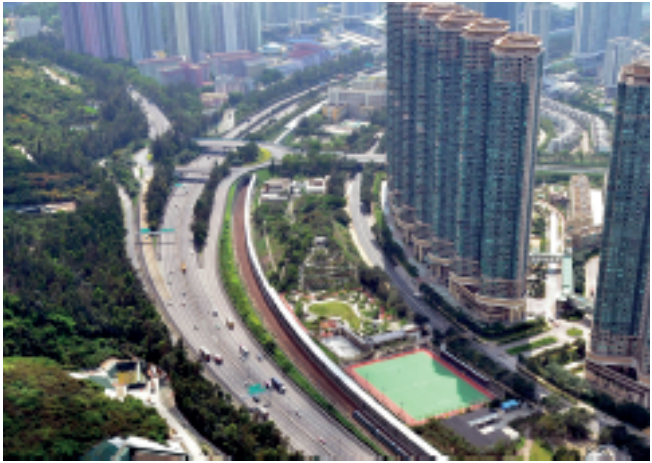

Figure 1. Cost and completion time information for 25 completed Hong Kong road projects was retrieved and analysed

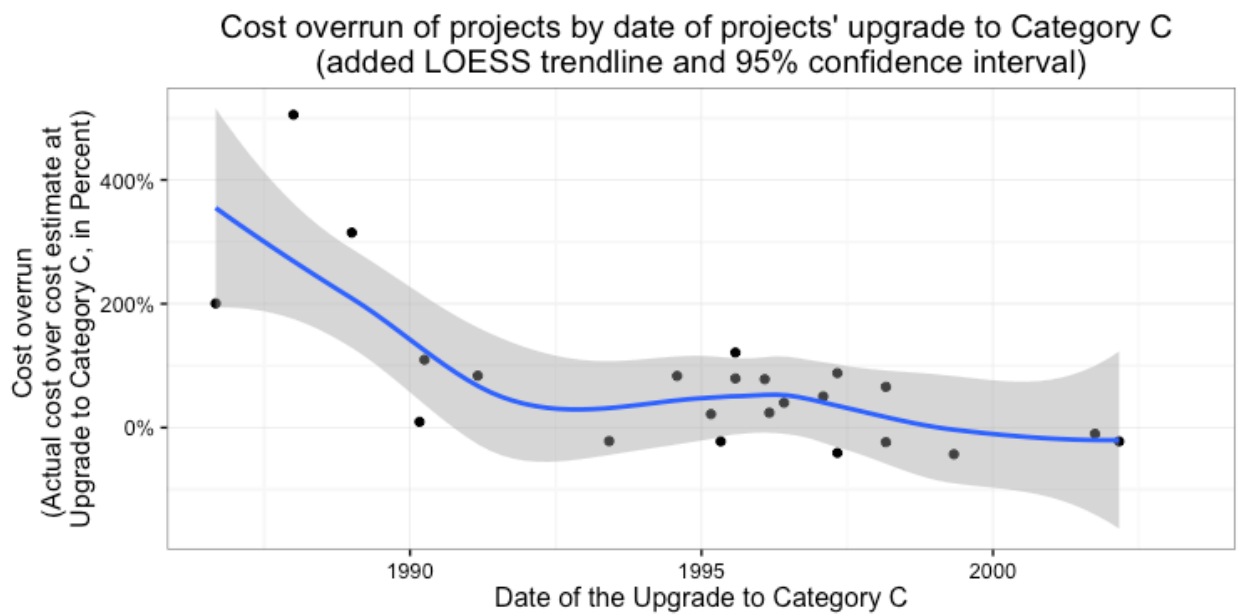

Figure 2. Cost overrun of projects by date of project's upgrade to category C (added Loess trendline and 95% confidence interval)

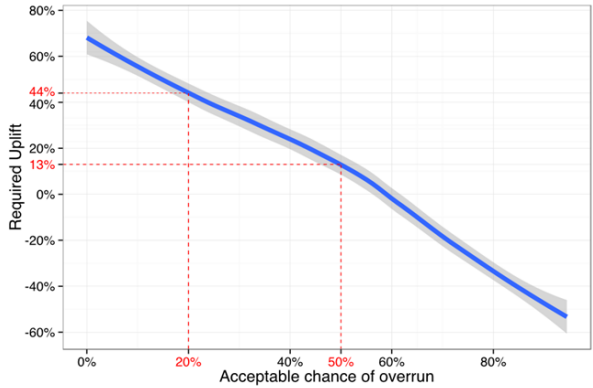

Figure 3. Required cost uplifts for different levels of the acceptable chance of a cost overrun for category C baseline cost estimates (added Loess smoother and 95% confidence interval)

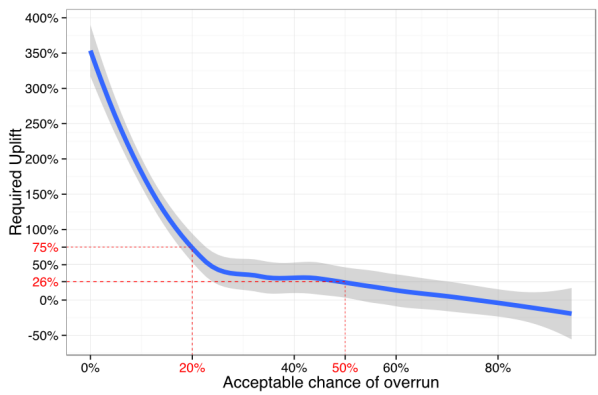

Figure 4. Required time uplifts for different levels of the acceptable chance of a delay for category C time-to-completion estimate (added Loess smoother and 95% confidence interval)

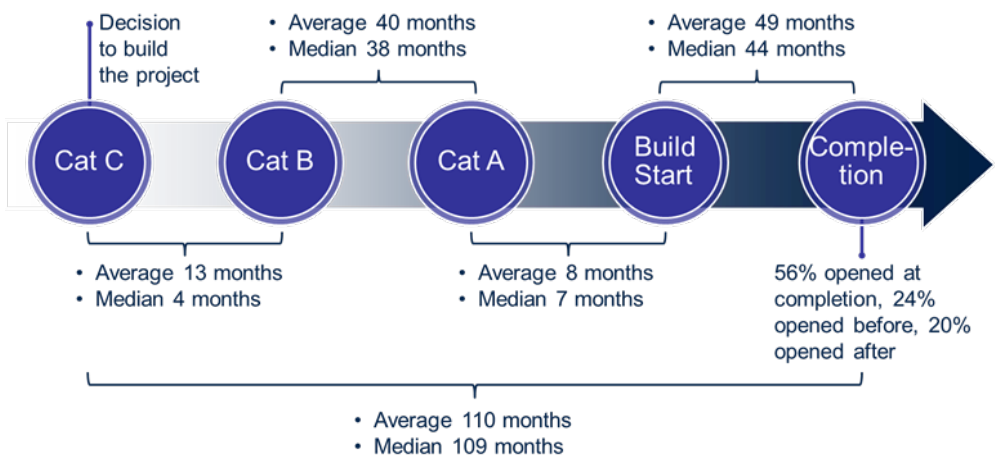

Figure 5. Summary of project duration